\documentclass[conference]{IEEEtran}
\IEEEoverridecommandlockouts

\usepackage{amsthm}
\usepackage{booktabs}
\usepackage{algorithm}
\usepackage{algorithmic}
\usepackage{tikz}
\usepackage{tikz-qtree}
\usepackage{color, colortbl}
\usepackage{xspace}

\usepackage{cite}
\usepackage{amsmath,amssymb,amsfonts}
\usepackage{algorithmic}
\usepackage{graphicx}
\usepackage{textcomp}
\usepackage{xcolor}
\usepackage{verbatim}

\def\BibTeX{{\rm B\kern-.05em{\sc i\kern-.025em b}\kern-.08em
    T\kern-.1667em\lower.7ex\hbox{E}\kern-.125emX}}
    
\usepackage{array}
\newcolumntype{L}[1]{>{\raggedright\let\newline\\\arraybackslash\hspace{0pt}}m{#1}}
\newcolumntype{C}[1]{>{\centering\let\newline\\\arraybackslash\hspace{0pt}}m{#1}}
\newcolumntype{R}[1]{>{\raggedleft\let\newline\\\arraybackslash\hspace{0pt}}m{#1}}    
    
\begin{document}

\title{Designing Explanations for\\ Group Recommender Systems}


\author{\IEEEauthorblockN{Alexander Felfernig}
\IEEEauthorblockA{\textit{Institute of Software Technology} \\
\textit{Graz University of Technology}\\
Graz, Austria \\
alexander.felfernig@ist.tugraz.at}
\and
\IEEEauthorblockN{Nava Tintarev}
\IEEEauthorblockA{\textit{Web Information Systems} \\
\textit{TU Delft}\\
Delft, the Netherlands \\
n.tintarev@tudelft.nl}
\and
\IEEEauthorblockN{Thi Ngoc Trang Tran}
\IEEEauthorblockA{\textit{Institute of Software Technology} \\
\textit{Graz University of Technology}\\
Graz, Austria \\
ttrang@ist.tugraz.at}
\and
\IEEEauthorblockN{Martin Stettinger}
\IEEEauthorblockA{\textit{Institute of Software Technology} \\
\textit{Graz University of Technology}\\
Graz, Austria \\
martin.stettinger@ist.tugraz.at}
}

\maketitle

\begin{abstract}
Explanations are used in recommender systems for  various reasons. Users have to be supported in making (high-quality) decisions more quickly. Developers of recommender systems want to convince users to purchase specific items. Users should better understand how the recommender system works and why a specific item has been recommended. Users should also develop a more in-depth understanding of the item domain. Consequently, explanations are designed in order to achieve specific \emph{goals} such as increasing the transparency of a recommendation or increasing a user's trust in the recommender system. In this paper, we provide an overview of existing research related to explanations in recommender systems, and specifically discuss aspects relevant to group recommendation scenarios. In this context, we present different ways of explaining and visualizing recommendations determined on the basis of preference aggregation strategies.
\end{abstract}

\begin{IEEEkeywords}
Recommender Systems, Group Recommender Systems, Explanations.
\end{IEEEkeywords}

\emph{Preprint, cite as}: A. Felfernig, N. Tintarev, T.N.T. Trang, and M. Stettinger. \emph{Explanations for Groups}. In A. Felfernig, L. Boratto, M. Stettinger, and M. Tkalcic (Eds.), Group Recommender Systems: An Introduction (pp. 105-126). SpringerBriefs in Electrical and Computer Engineering. Springer, 2018.


\section{Introduction}\label{sec:introduction:6}

\index{explanations} Explanations have been recognized as an important means to help users to evaluate recommendations, and make better decisions, but also to deliver persuasive messages to the user \cite{HerlockerKonstanRiedl2000,Tintarev2010}. Empirical studies show that users appreciate explanations of recommendations \cite{cramer2008effects,HerlockerKonstanRiedl2000}. Explanations can be regarded as \emph{a means to make something clear by giving a detailed description} \cite{TintarevMasthoff2012}. In the recommender systems context, Friedrich and Zanker \cite{FriedrichZanker2011} define explanations as \emph{information about recommendations} and as \emph{means to support objectives defined by the designer of a recommender system}. Explanations can be seen from two basic viewpoints \cite{Bilgic2005,Tintarev2016}: (1) the \emph{user's} (\emph{group member's}) and (2) the \emph{recommender provider's} point of view. \emph{Users of recommender systems} are in the need of additional information to be able to develop a better understanding of the recommended items. \emph{Developers of recommender systems} want to provide additional information to users for various reasons, for example, to convince the user to purchase an item, to increase a user's item domain knowledge (educational aspect), and to increase a user's \emph{trust in} and overall \emph{satisfaction with} the recommender system. Another objective is to make users more \emph{tolerant} with regard to  recommendations provided by the system. This is especially important for new users/items, otherwise a recommendation may be perceived as inappropriate. Solely providing the core functionality of recommender systems, i.e., showing a list of relevant items to users, could evoke the impression of interacting with a \emph{black box} with no transparency and no additional user-relevant information \cite{HerlockerKonstanRiedl2000,Tintarev2010}. Consequently, explanations are an important means to provide information related to recommendations, the recommendation process, and further objectives defined by the designer of a recommender system \cite{Chen2017,FriedrichZanker2011,Lamche2014,Sanchez2017,Verbert2013}. Visualizations of explanations can further improve the perceived quality of a recommender system \cite{Gansner2009,Tintarev2016,Verbert2013} -- where appropriate, examples of visualizations will be provided. 

\index{explanations}  \index{explanations ! single users}  \subsubsection*{Explanations in Single User Recommender Systems} In single user recommender systems, various efforts have already been undertaken to categorize explanations with regard to \emph{information sources used to generate explanations} and corresponding \emph{goals of explanations} \cite{FriedrichZanker2011,Gedikli2014,Nunes2017,Tintarev2009,Tintarev2010,TintarevMasthoff2015}. A categorization of different information sources that can be used for the explanation of recommendations is given, for example, in Friedrich and Zanker \cite{FriedrichZanker2011} where \emph{recommended items}, \emph{alternative items}, and the \emph{user model} are mentioned as three orthogonal \emph{information categories}. Potential goals of explanations are discussed a.o. in Tintarev and Masthoff \cite{Tintarev2010} and Jameson et al. \cite{Jameson2015}. Examples thereof are \emph{efficiency} (reducing the time needed to complete a choice task), \emph{persuasiveness} (exploiting explanations to change a user's choice behavior) \cite{Gkika2014}, \emph{effectiveness} (proactively helping the user to make higher-quality decisions), \emph{transparency} (reasons as to why an item has been recommended, i.e., answering why-questions), \emph{trust} (supporting a user in increasing her confidence in the recommender), \emph{scrutability} (providing ways to make the user profile manageable), \emph{satisfaction} (explanations focusing on aspects such as enjoyment and usability), and \emph{credibility} (assessed likelihood that a recommendation is accurate).  Bilgic and Mooney \cite{Bilgic2005} offer a differentiation between explanations that focus on (1) \emph{promotion}, i.e., convincing users to adopt recommendations, and (2) \emph{satisfaction}, i.e., to help users make more accurate decisions. 

Examples of verbal explanations for single user recommendations include phrases such as (1) '\emph{users who purchased item $x$ also purchased item $y$}', (2) '\emph{since you liked the book $x$, we recommend book $y$ from the same authors}', (3) '\emph{since you prefer taking sports photos, we recommend camera $y$ because it supports 10 pics/sec in full-frame resolution}', and (4) '\emph{item $y$ would be a good choice since it is similar to the already presented item $x$ and has the requested higher frame rate (pics/sec)}'. These example explanations are formulated based on  information collected and provided by the underlying recommendation approaches, i.e., (1) collaborative filtering, (2) content-based filtering, (3) constraint-based recommendation, and (4) critiquing-based recommendation -- see, for example, \cite{Chen2012,FelfernigGula2008,Gedikli2014,HerlockerKonstanRiedl2000}. These examples of explanations can be regarded as 'basic', since further information could be included. For instance, information related to competitor items and previous user purchases: '\emph{since you prefer taking sports photos, we recommend camera $y$ because it supports 10 pics/sec in full-frame resolution. $z$ would have been the other option but we propose $y$ since you preferred purchasing from provider $k$ in the past and $y$ is only a little bit more expensive than its competitors}'. 

Another type of explanation is the following: '\emph{no solution could be found -- if you increase the maximum acceptable price or decrease the minimum acceptable resolution, a corresponding solution can be identified.}' This explanation focuses on indicating options to find a way out of the 'no solution could be found' dilemma which  primarily occurs in the context of constraint-based recommendation scenarios \cite{felfernigburke08}. Another example is '\emph{item $y$ outperforms item $z$ in both, quality and price, whereas $x$ outperforms $z$ only in quality}'. This explanation does not focus on one item but supports the \emph{comparison} of different candidate items (in this case, $x$ and $y$). Importantly, it is directly related to the concept of \emph{asymmetric dominance} ($y$ outperforms $z$ two times whereas $x$ does this only once) which is a \emph{decoy effect} \cite{TeppanFelfernigIsak2011}. Explanations based on \emph{item comparisons} are mostly supported in critiquing-based \cite{Chen2012} and constraint-based recommendation \cite{FelfernigGula2006} which are both based on semantic recommendation knowledge. In critiquing-based recommendation, \emph{compound critiques} point out the relationship between the current reference item and the corresponding candidate items \cite{Mccarthy2004ThinkingPositively}. An example of a compound critique in the domain of \emph{digital cameras} is the following: \emph{on the basis of the current reference item $x$, you can take a look at cameras with a [lower price] and a [higher resolution] or at cameras with a [higher price] and a [higher optical zoom]}. An analysis of comparison interfaces in single user constraint-based recommendation is presented in \cite{FelfernigGula2006,felfernighotzbagleytiihonen2014}.

\index{explanations}  \index{explanations ! groups} \subsubsection*{Explanations in Group Recommender Systems}  The aforementioned explanation approaches focus on single users, and so, do not have to consider certain aspects of group decision making. Explanations for groups can have \emph{further goals} such as \emph{fairness} (taking into account, as far as possible, the preferences of all group members), \emph{consensus} (group members agree on the decision), and \emph{optimality} (a group makes an optimal or nearly-optimal decision\footnote{In contrast to single-user decision making, the exchange of decision-relevant knowledge among group members has to be  fostered \cite{Atas2017}.}). An important aspect in this context is that explanations show how the interests of individual group members are taken into account. This is not relevant in the context of single user recommender systems. Understanding the underlying process enables group members to evaluate the appropriateness of the way their preferences have to been taken into account by the group recommender system. Similar to explanations for single users, explanations for groups are shaped by the underlying recommendation algorithms. Explanations similar to those already mentioned can also be defined in a group context. For example, (1) '\emph{groups that like item $x$ also like item $y$}', (2) '\emph{since the group likes the film $x$, we also recommend film $y$ from the same director}', (3) '\emph{since the maximum camera \emph{price} accepted by group members is \emph{500} (defined by Paul) and the \emph{minimum accepted resolution} is \emph{18 mpix} (defined by Joe), we recommend $y$ which supports \emph{20 mpix} at a \emph{price} of \emph{459}.}', and (4) '\emph{item $x$ is a good choice since it supports a higher frame rate requested by all group members and is only a little bit more expensive}'. 

These examples show that the chosen preference aggregation approach has an impact on the explanation style. While \emph{aggregated predictions} \cite{felferniggrouprecalgoverview2018} include information about the individual preferences of group members (e.g., one group member specified the lowest maximum price of $500$) and thus support explanation goals such as \emph{fairness} and \emph{consensus}, \emph{aggregated models}-based approaches \cite{felferniggrouprecalgoverview2018} restrict explanations to the group level (e.g., groups that like $x$ also like $y$). More advanced (hybrid) explanations \cite{Kouki2017} can also be formulated in group recommendation scenarios, for example, '\emph{since all group members prefer sports photography, we recommend camera $y$ rather than camera $z$. It is only a little bit more expensive but has a higher usability which is important for group member \emph{Joe} who is a newbie in digital photography. Similar groups also preferred $y$}'. 

An example of an explanation in a situation where no solution could be found is: '\emph{no \emph{23 mpix} camera with a \emph{price} below \emph{250} could be found. Therefore we recommend  camera $y$ with \emph{20 mpix} and a \emph{price} of \emph{249} since price is the most important criterion for all group members.}' Finally, the following example shows how to take into account a group's social reality, for example, in terms of 'tactful' explanations \cite{Sanchez2017}: '\emph{Although your preference for item $y$ is not very high, your close friend Peter thinks it is an excellent choice}'. This example explanation is formulated on the level of \emph{aggregated predictions} and also takes into account social relationships among group members (e.g., neighborhoods in a social network).  On the level of \emph{aggregated models}, an explanation can be formulated as follows: '\emph{A majority thinks that it is a good choice. Some group members think that it is an excellent choice.}' (assuming the existence of at least some aggregated categorization of preferences such as \emph{number of likes}). Taking into account the individual preferences of group members helps to increase \emph{mutual awareness} among group members, and thus counteracts the natural tendency to focus on one's own favorite alternatives \cite{JamesonSmyth2007}. An approach to explaining the \emph{consequences of a given recommendation} is introduced by Jameson et al. \cite{51_Jameson2004}, where \emph{emotions} of individual group members with regard to a recommendation are visualized in terms of animated characters.  

We want to emphasize that \emph{explanations for groups} is a highly relevant research topic with a limited, but nevertheless direction-giving,  number of research results \cite{Ardissono2003,YuChen2011,Jameson2004,JamesonSmyth2007,Ntoutsi2012}. In the following, we sketch ways in which explanations for single-user recommendation scenarios can be adapted to groups. Following the idea of categorizing explanation types along the different recommendation approaches \cite{TintarevMasthoff2012,Vig2009}, we discuss explanations for groups in the context of \emph{collaborative-} and \emph{content-based filtering}, as well as \emph{constraint-} and \emph{critiquing-based recommendation}.

\section{Collaborative Filtering}\label{sec:cf:6}

\index{collaborative filtering}  \index{explanations ! collaborative filtering} A widely used example of explanations in collaborative filtering recommenders is '\emph{users who purchased item $x$ also purchased item $y$}'. Such explanations can be generated, for example, on the basis of \emph{association rule mining} which is often used as a model-based collaborative filtering approach \cite{Lin2002}. Herlocker et al. \cite{HerlockerKonstanRiedl2000} analyzed the role of explanations in collaborative filtering recommenders. They focused on the impact of different explanation styles on user acceptance of recommender systems. Explanations were mostly represented graphically. For example, a histogram of neighbors' ratings for the recommended item categorized ratings as 'good', 'neutral', or 'bad'. The outcome of their study was that rating histograms  are the most compelling way to explain rating data. Furthermore, \emph{simple graphs} were perceived as more compelling than more detailed explanations, i.e., \emph{simplicity of explanations is a key factor}.

An orthogonal approach to propose explanations for collaborative-filtering-based recommendations is presented by Chang et al. \cite{Chang2016}. Following the idea of generating recommendations based on knowledge from the crowd (see, e.g., \cite{Ulz2016}), the authors introduce the idea of asking crowd workers to provide feedback on explanations. Quality assurance is an issue but crowd-sourced explanations were considered high-quality. The authors mention \emph{longer explanation texts} and an \emph{increased number of references to item genres} as examples of indicators of high-quality explanations. An example of a question for crowd-sourcing in group recommendation scenarios is the following: '\emph{given this movie recommendation} (e.g., \emph{Guardians of the Galaxy}), \emph{which of the following are useful  explanations for a group of middle-aged persons?} \emph{Can be viewed by the whole family}; \emph{Includes plenty of songs from the 70ies};  \emph{Best movie we have ever seen}'. This way, crowd knowledge can be exploited to better figure out which kinds of explanations are useful in which context and which ones might be particularly well-received by specific groups (in this case, middle-aged persons). A similar approach can be used to figure out relevant explanations in other recommendation approaches, i.e., \emph{which tags to use for an explanation}? (content-based filtering), \emph{which requirements to relax}? (constraint-based recommendation), and \emph{which critiques to propose to the user}? (critiquing-based recommendation).

As mentioned by Bilgic and Mooney \cite{Bilgic2005}, a goal of the explanations introduced in Herlocker et al. \cite{HerlockerKonstanRiedl2000} is to promote items but not to provide more insights as to why the items have been recommended, i.e., not to provide satisfaction-oriented explanations that might help users to make more accurate decisions. There are different ways to move the explanation focus towards more informative explanations. As proposed in \cite{Bilgic2005} (for single user recommenders), a collaborative-filtering-based explanation can be extended by providing information on items that had a major influence on the determination of the proposed recommendation. Removing the \emph{most influential items} (already rated by group members) from the set of rated items triggers the most significant difference in terms of recommended item ratings. Similar approaches can be used to determine the most influencing items in other recommender types \cite{Bilgic2005,Symeonidis2008}.

\index{collaborative filtering}  \index{explanations ! collaborative filtering} \subsubsection*{Collaborative Filtering Explanations for Groups} An example of basic explanations in group-based collaborative filtering is included in \textsc{PolyLens}, where the predicted rating for each group member and for the group as a whole is shown \cite{Connor2001}. Some simple examples of how to provide explanations in the context of group-based collaborative filtering scenarios are provided in Tables \ref{tab:CFExplanationsPromotionAggregatedPredictions:6} and \ref{tab:CFExplanationsPromotionAggregatedModels:6}. Both examples represent variants of the explanation  approaches introduced by Herlocker et al. \cite{HerlockerKonstanRiedl2000}. Table \ref{tab:CFExplanationsPromotionAggregatedPredictions:6} depicts an example of an explanation that is based on the preferences (ratings) of the nearest neighbors ($NN = \bigcup \{n_{ij}\}$) of the group members $u_i$ (for simplicity, we assume the availability of a complete set of rating data). For each recommended item $t_i$, the corresponding frequency distribution of the ratings of the nearest neighbors of individual group members is shown. Note that $NN$ can represent users who are in the intersection of users who rated this item ($\{n_{11},n_{12}, ...\} \cap ... \cap  \{n_{m1},n_{mk}, ...\}$). Alternatively, $NN$ can represent the  users in the union of nearest neighbors ($\{n_{11},n_{12}, ...\} \cup ... \cup  \{n_{m1},n_{mk}, ...\}$). A related explanation can be '\emph{users similar to members of this group rated item $t$ as follows}'.

\begin{table*}[ht!]
\center
\begin{tabular}{|C{0.75cm}|C{0.75cm}|C{0.75cm}|C{0.75cm}|C{0.75cm}|C{0.75cm}|C{0.75cm}|C{1.5cm}|C{1.5cm}|C{1.5cm}|} 
\hline
rec. item $t_i$  	&  \multicolumn{6}{  c  |}{ratings of nearest neighbors $n_{ij} \in NN$} 	&  \multicolumn{3}{  c  |}{explanation} \\ \cline{2-10}
  		&  \multicolumn{2}{  c  |}{$u_1$}&  \multicolumn{2}{  c  |}{$u_2$}	&  \multicolumn{2}{  c  |}{$u_3$} 	&  bad ~~~~[$0-2$]	&  neutral [$>2-3.5$] 	&  good [$>3.5-5$] \\ \cline{2-7}
        &  $nn_{11}$ 	&	$nn_{12}$	&  $nn_{21}$	& $nn_{22}$	&  $nn_{31}$	& $nn_{32}$					&		&   		&   	\\ \hline
$t_1$ 	&  4.2			&  4.9			& 4.3			&	3.5		&   3.2			&	4.8						&   0	&	2		& 4  	\\ \hline 
$t_2$   &  3.5			&  2.2			& 2.7			&	3.2		&   2.9			&	3.6						&  	0	&	5		& 1  	\\ \hline 
$t_3$  	&  3.8			&  3.1			& 3.7			&	2.8		&   3.4			&	2.6						&  	0	&	4		& 2  	\\ \hline 
$t_4$  	&  4.3			&  4.9			& 4.4			&	4.5		&   4.0			&	4.0						&  	0	&	0		& 6  	\\ \hline  
$t_5$  	&  3.7			&  3.9			& 3.2			&	3.5		&   3.6			&	2.9						&  	0	&	3		& 3  	\\ \hline 
\end{tabular} \vspace{0.2cm}
\caption{Collaborative filtering explanations for \emph{aggregated predictions}, i.e., explanations based on information about the preferences (ratings) of nearest neighbors ($n_{ij}$) of individual group members $u_i$.}
\label{tab:CFExplanationsPromotionAggregatedPredictions:6}
\end{table*}

Table \ref{tab:CFExplanationsPromotionAggregatedModels:6} depicts an example of an explanation that is based on the preferences of neighborhood groups $gp_j$ of the current group $gp$. We assume that ratings are only available in an aggregated fashion (ratings of individual users are not available, e.g., for privacy reasons). In this context, the frequency distribution of the ratings of the nearest neighbor groups is shown for each item $t_i$. An explanation can contain the following text: '\emph{groups similar to the current group rated item $t$ as follows}'.

\begin{table*}[ht!]
\center
\begin{tabular}{|C{0.75cm}|C{0.75cm}|C{0.75cm}|C{0.75cm}|C{0.75cm}|C{1.5cm}|C{1.5cm}|C{1.5cm}|} 
\hline
rec. item  	&   \multicolumn{4}{  c  |}{ratings of NN groups ($gp_{j}$)} 	&  \multicolumn{3}{  c  |}{explanation} \\ \cline{2-8}
  		&  $gp_1$	&  $gp_2$ 	&  $gp_3$ 	&  $gp_4$ 	&  bad ~~~~[$0-2$]&  neutral [$>2-3.5$] 	&  good [$>3.5-5$] \\ \hline
$t_1$ 	&  4.2		&  4.9		& 4.3		&	 3.5	&	 0	&	1		&   3	\\ \hline 
$t_2$  	&  1.2		&  2.9		& 3.1		&	 1.8	&	 2	&	2		&   0	\\ \hline 
$t_3$  	&  3.5		&  3.8		& 2.9		&	 3.3	&	 0	&	3		&   1	\\ \hline 
$t_4$  	&  4.9		&  4.8		& 4.1		&	 4.4	&	 0	&	0		&   4	\\ \hline 
$t_5$  	&  3.7		&  3.3		& 2.4		&	 3.9	&	 0	&	2		&   2	\\ \hline 
\end{tabular} \vspace{0.2cm}
\caption{Collaborative filtering explanations for \emph{aggregated models}, i.e., explanations are based on the aggregated preferences of individual group members.}
\label{tab:CFExplanationsPromotionAggregatedModels:6}
\end{table*}

In the given examples, explanations refer to ratings but do not take into account aggregation functions. Ntoutsi et al. \cite{Ntoutsi2012} present an approach to explain the  aggregation functions in aggregated-prediction-based collaborative filtering. For example, the application of \emph{Least Misery} (LMS) triggers explanations of type '\emph{item $y$ has a group score of 2.9 due to the (lowest) rating determined for user $a$}'. A more 'group-oriented' explanation is '\emph{item $y$ is recommended because it avoids misery within the group}'. When using \emph{Most Pleasure} (MPL), the corresponding explanation would be '\emph{item $y$ has a group score of 4.8 due to the (highest) rating determined for user $b$}'. Finally, when using  \emph{Average} (AVG), explanations of type '\emph{item $y$ is most similar to the ratings of users $a,b$, and $c$}' are provided. Similar explanations can be generated for content-, constraint-, and critiquing-based recommendations. Although initial approaches have already been proposed, different ways to explain group recommendations depending on the used aggregation function(s) are an issue for future research.

\index{collaborative filtering}  \index{visualization ! collaborative filtering} \subsubsection*{Visualization of Collaborative Filtering Explanations for Groups} 
\begin{figure}[ht!]
\centering
\includegraphics[scale=.4]{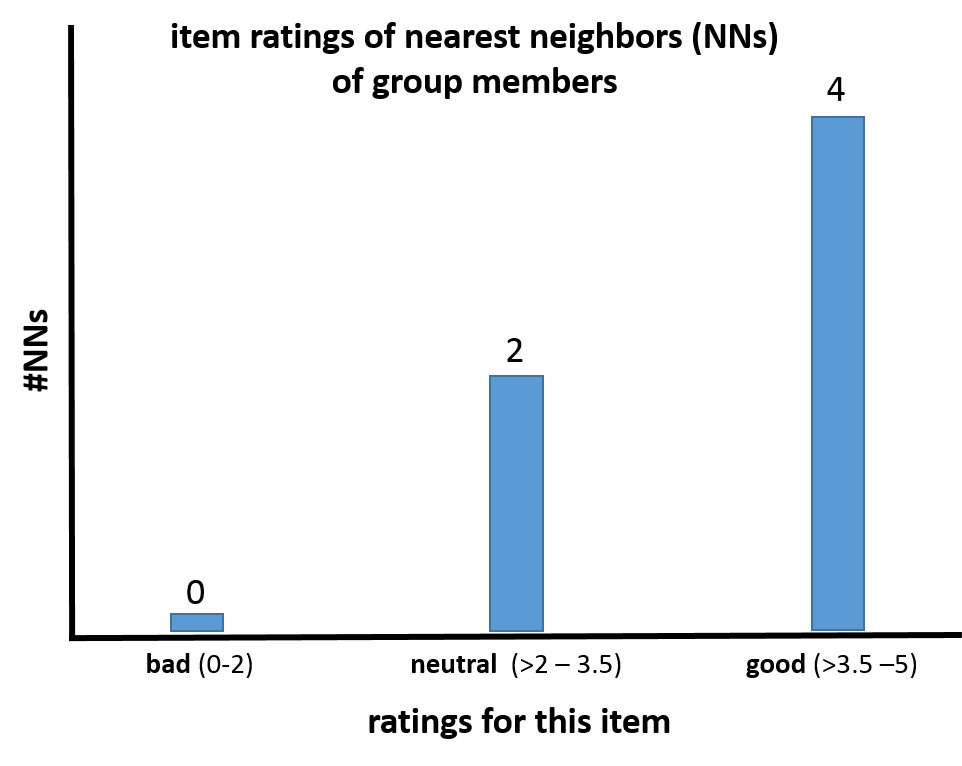}
\caption{Graphical representation of the explanation data contained in Table \ref{tab:CFExplanationsPromotionAggregatedPredictions:6}.}
\label{fig:VisNNsCF:6}       
\end{figure}

\begin{figure}[ht!]
\centering
\includegraphics[scale=.4]{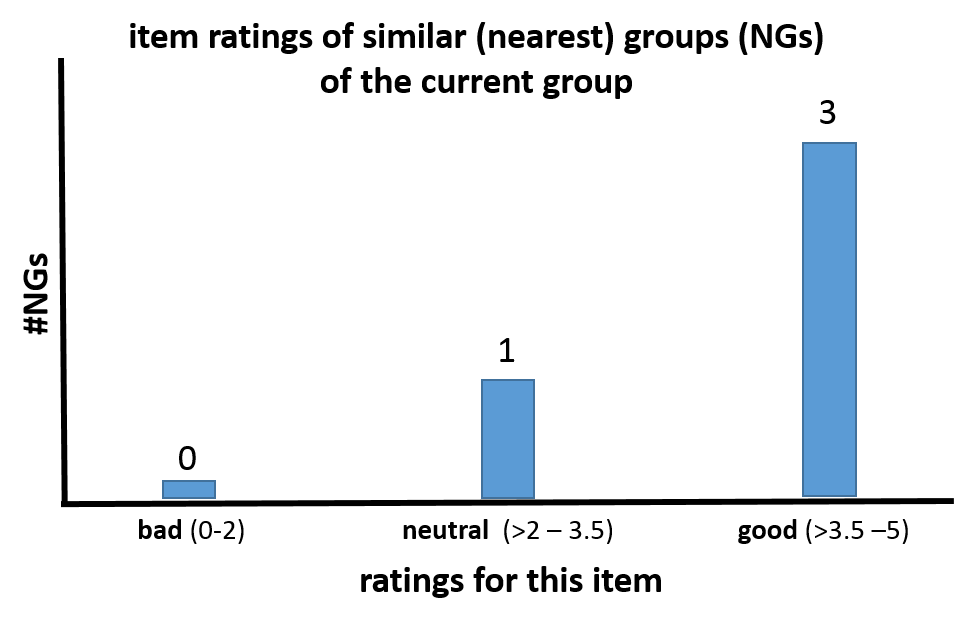}
\caption{Graphical representation of the explanation data contained in Table \ref{tab:CFExplanationsPromotionAggregatedModels:6}.}
\label{fig:VisNGsCF:6}       
\end{figure}

\begin{figure}[ht!]
\centering
\includegraphics[scale=.5]{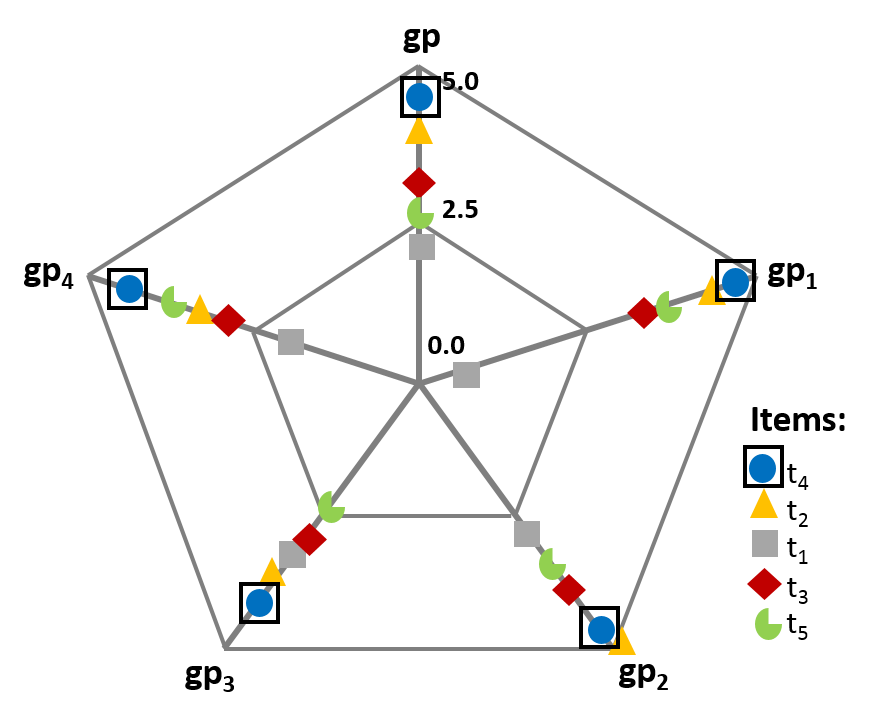}
\caption{Spider diagram for explaining aggregated models based collaborative filtering recommendations: ratings of nearest neighbor groups $gp_1, .., gp_4$ of $gp$ for the recommended item $t_4$. This representation is a variant of consensus-based interfaces discussed in \cite{Mahyar2017}.}
\label{fig:SpiderDiagramCF:6}       
\end{figure}

There are different ways to visualize a recommendation determined using collaborative filtering \cite{HerlockerKonstanRiedl2000}. The frequency distributions introduced and evaluated by Herlocker et al. \cite{HerlockerKonstanRiedl2000} can also be applied in the context of group recommendation scenarios. An example thereof is given in Figure \ref{fig:VisNNsCF:6}, where the explanation information contained in Table \ref{tab:CFExplanationsPromotionAggregatedPredictions:6} is represented graphically. Figure \ref{fig:VisNGsCF:6} depicts a similar example where an item-specific evaluation of nearest (most similar) groups is shown in terms of a frequency distribution. Alternatively, \emph{spider diagrams} can be applied to visualize the preferences of nearest neighbors. An example is depicted in Figure \ref{fig:SpiderDiagramCF:6}. This type of representation is based on the idea of consensus-based approaches to visualize the current status of a group decision process \cite{Mahyar2017,Palomares2014}.

\section{Content-based Filtering}\label{sec:cbf:6}

\index{content-based filtering}  \index{explanations ! content-based filtering}  The basis for determining recommendations in content-based filtering is the similarity between item  descriptions and keywords (categories) stored in a user profile. Since the importance of keywords can differ among group members, it is important to identify those which are relevant for all group members \cite{Lieberman1999}. Explanations are based on the analysis of item-related content. Examples of verbal explanations  in content-based filtering  are given in \cite{Bilgic2005}. The authors show that keyword-style explanations can increase both the  perceived trustworthiness and the transparency of recommendations. Such explanations primarily represent occurrence statistics of keywords in item descriptions (see also \cite{cramer2008effects}). Gedikli et al. \cite{Gedikli2014} compare different approaches to representing explanations in content-based filtering scenarios, and show that tag-cloud-based graphical representations outperform verbal approaches. 

\subsubsection*{Content-based Filtering Explanations for Groups} A simple example of content-based filtering explanations for groups  is depicted in Table \ref{tab:CBFExplanationsPromotionAggregatedModels:6}.

\begin{table*}[ht!]
\center
\begin{tabular}{|C{1.25cm}|C{0.65cm}|C{0.65cm}|C{0.65cm}|C{0.65cm}|C{0.65cm}|C{0.65cm}|C{0.65cm}|C{1.0cm}|C{1.0cm}|C{1.0cm}|C{1.0cm}|} 
\hline
category  	& \multicolumn{3}{  c  |}{userweights} 	&  \multicolumn{4}{  c  |}{itemweights} 	&  \multicolumn{4}{  c  |}{explanation-relevance} 		\\ \cline{2-12}
        	& $u_1$ & $u_2$	& $u_3$ &  $t_1$	&  $t_2$ 	&  $t_3$ 	&  $t_4$ 	&  $t_1$		&  $t_2$ 		&  $t_3$ 		&  $t_4$ 		\\ \hline
$cat_1$ 	& 0.05 	& 0.1 & 0.15	& 0.1		&  0.1		&  0.2		&    0.3	&   0.01			&  0.01			&  0.02 		&  0.03 		\\ \hline 
$cat_2$  	& 0.3 	& 0.4 & 0.5		& 0.7		&  0.2		& 0.2		&	 0.0	&  0.28$\surd$	&  0.08$\surd$ 	&  0.08 		&  0.0			\\ \hline 
$cat_3$  	& 0.15  & 0.25 & 0.2	& 0.1		&  0.4		& 0.2		&	 0.3	&  0.02			&  0.08$\surd$ 	&  0.04 		&  0.06$\surd$	\\ \hline 
$cat_4$  	& 0.4   & 0.3  & 0.2	& 0.1		&  0.2		& 0.3		&	 0.1	&  0.03			&  0.06 		&  0.09 $\surd$ &  0.03			\\ \hline 
\end{tabular} \vspace{0.2cm}
\caption{Content-based filtering explanations for \emph{aggregated predictions}. The most explanation-relevant categories for an item $t_k$ are marked with $\surd$.}
\label{tab:CBFExplanationsPromotionAggregatedModels:6}
\end{table*}

Item categories $cat_j$ have a user-specific weight (derived, for example, from the category weights of individual user profiles where user $u_i$ is a member of group $G$).  To determine the \emph{explanation relevance} ($er$) of individual categories, these weights are combined with item-individual weights ($iw$)  (see Formula \ref{explanationrelevance}). 

\begin{equation} \label{explanationrelevance}
er(cat_j, t_k) = \frac{\Sigma_{u_i \in G} userweight(u_i,cat_j) \times iw(t_k,cat_j)}{|G|}
\end{equation}

The higher the explanation-relevance of a category, the higher the category will be ranked in a list shown to the group (members). A verbal explanation related to item $t_1$ (Table  \ref{tab:CBFExplanationsPromotionAggregatedModels:6}) can be of the form '\emph{item $t_1$ is recommended since each group member is interested in category $cat_2$}'. If the preference information of individual group members is not available (e.g., for privacy reasons), this explanation would be formulated as '\emph{item $t_1$ is recommended since the group as a whole is interested in category $cat_2$}'. Also, more than one category can be used in such an explanation.  As mentioned, category- or keyword-based explanations can also be extended with information about the most influential items \cite{Bilgic2005}. This can be achieved by determining those items that trigger the most significant change in item rating predictions (if not taken into account by the recommendation algorithm).

An approach to explaining recommendations on the basis of tags is presented in Vig et al. \cite{Vig2009}. \emph{Tagsplanations} (explanations based on user community tags) are introduced to explain recommendations. In this context, \emph{tag relevance} is defined as the \index{Pearson correlation} Pearson Correlation between item ratings and corresponding tag preference values. \emph{Tag preference} is the relationship between the number of times a specific tag has been applied to an item compared to the total number of tags applied to the item (weighted with corresponding item ratings). In a study with \textsc{MovieLens} \cite{Miller2004} users, the authors show that both tag relevance and tag preference help to achieve the explanation goals of \emph{justification} (why has an item been recommended) and \emph{effectiveness} (better decisions are made). Similar to the example shown in Table \ref{tab:CBFExplanationsPromotionAggregatedModels:6}, explanation-relevance (in this case tag relevance) is used to order a list of explanatory tags \cite{Vig2009}. 

An \emph{opinion mining} approach to generating explanations is introduced by Muhammad et al. \cite{Muhammad2016}. In the context of opinion mining, features are extracted from item reviews \cite{Dong2013} and then associated with corresponding sentiment scores. Features and corresponding sentiments are then used to generate explanations related to the \emph{pros} and \emph{cons} of specific items. Features are sorted into \emph{pro} or \emph{con} according to whether their values are above or below a predetermined threshold. If we assume, for example, a threshold of $0.4$, all item features with an explanation relevance $\geq 0.4$ are regarded as pros, the others are regarded as cons. Formula \ref{explanationrelevanceopinionmining} represents an approach to determine the explanation-relevance ($er$) of a specific feature $f_i$ where \emph{sentiment} represents a group preference with regard to a specific feature and \emph{item-sentiment} represents the support of the feature by the item $t_j$. 

\begin{equation} \label{explanationrelevanceopinionmining}
er(f_i) = sentiment(f_i) \times item\text{-}sentiment(t_j,f_i)
\end{equation}

Opinion mining approaches to explanations can also be extended to groups. An  example of applying Formula \ref{explanationrelevanceopinionmining} in the context of group recommender systems is given in Table \ref{tab:CBFExplanationsPromotionAggregatedPredictionsOpinions:6}. 

\begin{table*}[ht!]
\center
\begin{tabular}{|C{1.5cm}|C{1.5cm}|C{0.75cm}|C{0.75cm}|C{0.75cm}|C{0.75cm}|C{1.0cm}|C{1.0cm}|C{1.0cm}|C{1.0cm}|} 
\hline
 \multicolumn{2}{|  c  |}{group profile ($gp$)} 	&  \multicolumn{4}{  c  |}{item-sentiments} 	&  \multicolumn{4}{  c  |}{explanation-relevance} \\ \cline{1-10}
feature & sentiment	&  $t_1$	&  $t_2$ 	& $t_3$ 	&  $t_4$ 		&  $t_1$		&  $t_2$ 	&  $t_3$ 	&  $t_4$ 	\\ \hline
$f_1$ 	& 0.10		&  0.19		&  0.23		& 0.35		&	0.68		&  0.019		&  0.023 	&  0.035 	&  0.068		\\ \hline 
$f_2$  	& 0.76		&  0.61		&  0.52		& 0.47		&	0.52		&  0.46			&  0.40 	&  0.36 	&  0.40		\\ \hline 
$f_3$  	& 0.21		&  0.47		&  0.43		& 0.21		&	0.31		&  0.10			&  0.09 	&  0.04 	&  0.07		\\ \hline 
$f_4$  	& 0.82		&  0.92		&  0.76		& 0.49		&	0.77		&  0.75$\surd$	&  0.62$\surd$ 	&  0.40 $\surd$ &  0.63$\surd$		\\ \hline 
\end{tabular} \vspace{0.2cm}
\caption{Opinion mining based explanations for \emph{aggregated models}. Features $f_i$ with the highest explanation-relevance are marked with $\surd$.}
\label{tab:CBFExplanationsPromotionAggregatedPredictionsOpinions:6}
\end{table*}

This example sketches the generation of explanations in \emph{aggregated models} scenarios. When determining explanations in the context of \emph{aggregated predictions}, explanation relevance could be determined for each individual user and then aggregated using an aggregation function such as \emph{Average} ($AVG$)                                                                   to select explanations considered most relevant for the group.

\index{content-based filtering}  \index{visualization ! content-based filtering} \subsubsection*{Visualization of Content-based Filtering Explanations for Groups} 
\begin{figure}[ht!]
\centering
\includegraphics[scale=.48]{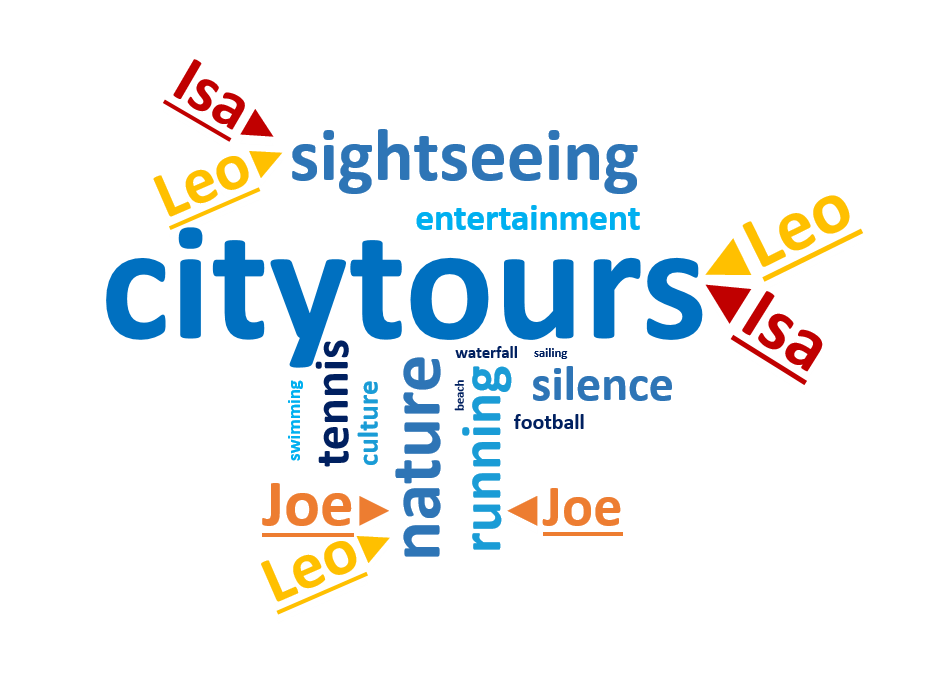}
\caption{Tag-cloud representation used to show the relevance of tags with regard to a specific item extended with preference information related to group members (\emph{Isa}, \emph{Joe}, and \emph{Leo}).}
\label{fig:TagCloudCBF:6}       
\end{figure}

An alternative to list-based representations of explanations is mentioned, for example, in Gedikli et al. \cite{Gedikli2014}, where content-based explanations are visualized in the form of \emph{tag-clouds}. An example of a tag-cloud-based explanation in the context of group recommendation is depicted in Figure \ref{fig:TagCloudCBF:6}. The used tags are related to the travel domain. In this scenario, the tag-cloud represents an explanation based on the \emph{aggregated preferences} of individual group members. For example, \emph{Leo} and \emph{Isa} like city tours. One can imagine other visual encodings in terms of shape, textures, and highlightings \cite{Knutov2009}. Tag relevance can be determined on the basis of a tag relevance estimator similar to Formula \ref{explanationrelevance}. 


\section{Constraint-based Recommendation}\label{sec:cbr:6}

\index{constraint-based recommendation}  \index{explanations ! constraint-based recommendation} Constraint-based recommender systems are built upon deep knowledge about items and their corresponding recommendation rules (constraints). This information serves as a basis for explaining item recommendations by analyzing reasoning steps that led to the derivation of solutions (items) \cite{Friedrich2004}. Such explanations follow the tradition of AI-based expert systems \cite{Buchanan1984,FriedrichZanker2011}. On the one hand, explanations are used to answer \emph{how}-questions, i.e., questions related to the reasons behind a recommendation. A corresponding analysis is provided, for example, by Felfernig et al. \cite{FelfernigGula2006}. \emph{How} questions are answered in terms of showing the relationship between defined user requirements $req_i$ and the recommended items. An example of such an explanation is '\emph{item $y$ is recommended, since you specified the upper price limit with $500$ and you preferred light-weight cameras}' (for  details see \cite{FelfernigGula2006,Friedrich2004}). Besides answering \emph{how} questions, constraint-based recommenders help to answer \emph{why} and \emph{why not} questions. Explanations for the first type are used to provide insights to the user as to why certain questions have to be answered, whereas explanations for \emph{why not} questions help a user to escape from the \emph{no solution could be found} dilemma \cite{FelfernigSchubertFriedrichMandlMairitschTeppan2009}.  Felfernig et al. \cite{FelfernigGula2006} show that such explanations can help to increase a user's trust in the recommender application. Furthermore, explanations related to  \emph{why not} questions can increase the perception of item domain knowledge.

\subsubsection*{Explanations in Constraint-based Recommendation for Groups} Formula \ref{explanationrelevanceconstraintbasedrecommendation} represents a simple example of an approach to determine the explanation-relevance ($er$) of user requirements in constraint-based recommendation scenarios for groups. A related example is depicted in Table \ref{tab:CBRExplanationsPromotionAggregatedPredicitons:6}. The assumption is that all group members have already agreed on the set of requirements $\bigcup req_j$ and each group member has also specified  his/her preference in terms of an importance value. An explanation that can be provided to a group in such a context is '\emph{requirement $req_3$ is considered important by the whole group}'. 

\begin{equation} \label{explanationrelevanceconstraintbasedrecommendation}
er(req_j) = \frac{\Sigma_{u_i \in G} importance(req_j, u_i)}{|G|}
\end{equation}

\begin{table*}[ht!]
\center
\begin{tabular}{|C{1.5cm}|C{0.5cm}|C{0.5cm}|C{0.5cm}|C{1.5cm}|} 
\hline
requirement & \multicolumn{3}{  c  |}{importance}	&  explanation relevance 	\\ \cline{2-4}
  			& $u_1$	&  $u_2$ & $u_3$ 				&  							\\ \hline
$req_1$ 	& 0.2	&   0.3	 & 0.4					&  0.3 						\\ \hline 
$req_2$  	& 0.5	&   0.4	 & 0.1					&  0.33 					\\ \hline 
$req_3$  	& 0.3	&   0.3	 & 0.5					&  0.37$\surd$ 				\\ \hline 
\end{tabular} \vspace{0.2cm}
\caption{Explanation relevance of requirements in constraint-based recommendation (\emph{aggregated models}). The most relevant requirement is marked with $\surd$.}
\label{tab:CBRExplanationsPromotionAggregatedPredicitons:6}
\end{table*}

The example explanation shown in Table \ref{tab:CBRExplanationsPromotionAggregatedPredicitons:6} does not take into account causal relationships between requirements and items \cite{Friedrich2004}. For example, if a group agrees that the \emph{price} of a camera has to be below $1,000$ and every camera fulfills this criteria, the price requirement does not filter out items from the itemset, so there is no causal relationship between a recommendation subset of a given itemset and the price requirement. 

\subsubsection*{Combining Constraints and Utilities} Constraint-based recommendation is often \emph{combined with} an additional mechanism that supports the ranking of candidate items. An example thereof is Multi-Attribute Utility Theory (MAUT) \cite{WinterfeldtEdwards1986} that supports the evaluation of items in terms of a set of \emph{interest dimensions} which can be interpreted as generic requirements. For example, in the digital camera domain, \emph{output quality} is an interest dimension that is related to user requirements such as \emph{resolution} and \emph{sensor size}. Group members specify their preferences with regard to the importance of the  interest dimensions $dim_i$.  Furthermore, items $t_j$ have different contributions with regard to these dimensions (see Table \ref{tab:CBRExplanationsPromotionAggregatedPredictionsMAUT:6}).

\begin{table*}[ht!]
\center
\begin{tabular}{|C{1.5cm}|C{0.5cm}|C{0.5cm}|C{0.5cm}|C{0.5cm}|C{0.5cm}|C{0.5cm}|C{1.0cm}|C{1.0cm}|C{1.0cm}|} 
\hline
dimension  	& \multicolumn{3}{  c  |}{importance}	&  \multicolumn{3}{  c  |}{contribution} 	&  \multicolumn{3}{  c  |}{explanation relevance} \\ \cline{2-10}
  			& $u_1$	&  $u_2$ & $u_3$ 				& $t_1$	&  $t_2$ 	&  $t_3$ 			&  $t_1$	&  $t_2$ 		&  $t_3$ 		\\ \hline
$dim_1$ 	& 0.1	&   0.3	 & 0.1					&  0.3	&  0.3		& 0.1			&  0.05			&  0.05 		&  0.02 		\\ \hline 
$dim_2$  	& 0.6	&   0.5	 & 0.3					&  0.3	&  0.5		& 0.6			&  0.14			&  0.23$\surd$ 	&  0.28$\surd$ 	\\ \hline 
$dim_3$  	& 0.3	&   0.2	 & 0.6					&  0.4	&  0.2		& 0.2			&  0.15$\surd$	&  0.07 		&  0.07 		\\ \hline 
\end{tabular} \vspace{0.2cm}
\caption{Explanation relevance of interest dimensions in utility-based recommendation (\emph{aggregated predictions}). The most relevant dimension is marked with $\surd$.}
\label{tab:CBRExplanationsPromotionAggregatedPredictionsMAUT:6}
\end{table*}

Similar to content-based filtering, the item-specific \emph{explanation relevance} ($er$) of individual interest dimensions can be determined on the basis of Formula \ref{explanationrelevanceconstraints} where $imp$ represents the user-specific importance of an interest dimension $dim_i$ and $con$ the contribution of an item to $dim_i$. 

\begin{equation} \label{explanationrelevanceconstraints}
er(dim_i,t_j) = \frac{\Sigma_{u_k \in G} (imp(u_k,dim_i) \times con(t_j,dim_i))}{|G|}
\end{equation}

Following this approach, \cite{Carenini2006,felfernig2008_persuasion,Symeonidis2008,Teze2015} show how to apply utility-based approaches to the selection of evaluative arguments\footnote{In line with Jameson and Smyth \cite{JamesonSmyth2007}, we interpret arguments as elementary parts of explanations.}, i.e.,  arguments with the highest relevance. In this context, arguments take over the role of the previously-mentioned interest dimensions. Such an approach is provided in the \textsc{Intrigue} system \cite{Ardissono2003}, where recommended travel destinations are explained to groups, and arguments are chosen depending on their utility for individual group members or subgroups.

An example of an argument (as an elementary component of an explanation) for a car recommended by a constraint-based recommender is \emph{'very energy-efficient'}, where \emph{energy-efficiency} can be regarded as an interest dimension. The contribution of an item to this interest dimension is high if, for example, the fuel consumption of a car is low. If a customer is interested in energy-efficient cars and a car is energy efficient, the corresponding argument will be included in the explanation (see the example in Table \ref{tab:CBRExplanationsPromotionAggregatedPredictionsMAUT:6}). An example explanation from another domain (financial services) is the following: '\emph{financial service $t_1$ is recommended since all group members strongly prefer low-risk investments}'. Examples of interest dimensions used in this context are \emph{risk}, \emph{availability}, and \emph{profit}.

\subsubsection*{Consensus in Group Decisions} 

\index{explanations ! consensus} \index{consensus} Situations can occur where the preferences of individual group members become inconsistent \cite{FelfernigGroupBasedConfiguration2016,FelfernigSchubertZehentner2012,Mahyar2017}. In the context of group recommendation scenarios, consensus is defined in terms of disagreement between individual group members regarding item evaluations (ratings) \cite{AmerYahia2009}. To provide a basis for establishing consensus, such situations have to be explained and visualized \cite{Jameson2004,Mahyar2017}. In this context, diagnosis methods \cite{FelfernigSchubertZehentner2012} can help to determine repair actions that propose changes to the current set of requirements (preferences) such that a recommendation can be identified. Such repairs are able to take into account the individual preferences of group members \cite{FelfernigGroupBasedConfiguration2016}. The potential of aggregation functions  to foster consensus in group decision making is discussed in Salamo et al. \cite{Salamo2012}. Concepts to take into account consensus in group decision making are also presented in \cite{AmerYahia2009,Castro2017,Castroetal2015Consensus}. In scenarios such as software requirements engineering \cite{Ninaus2014}, there are often  misconceptions regarding the evaluation/selection of a specific requirement. For example, there could be misconceptions regarding the assignment of a requirement to a software release. An explanation in such contexts indicates possible changes of requirements (assignments) that help to restore consistency. In group-based settings, such  repair-related explanations help group members understand the constraints of other group members and decide in which way their own requirements should be adapted.

\index{explanations ! user-generated} \subsubsection*{User-generated Explanations}  User-generated explanations are defined by a group member (typically, the creator of a decision task) to explain, for example, why a specific alternative has been selected. The impact of user-generated explanations in constraint-based group recommendation scenarios was analyzed  by Stettinger et. al \cite{Stettinger2015UMAP}. The creator of a decision task (prioritization decisions in the context of software requirements engineering) had to explain the decision outcome verbally. In groups where such explanations were provided, this contributed to an increased  satisfaction with the final decision and an increased perceived degree of group decision support quality \cite{Stettinger2015UMAP}. User-generated explanations are not limited to constraint-based recommendation.  For example, crowd-sourcing based approaches are based on the similar idea of collecting explanations directly from users.

\index{explanations ! fairness} \index{fairness} \subsubsection*{Fairness Aspects in Groups} \emph{Fair recommendations in group settings} can be  characterized as \emph{recommendations without favoritism or discrimination towards specific group members}. The perceived importance of fairness, depending on the underlying item domain, has been analyzed in \cite{AtasIEA2017}. An outcome of this study is that in high-involvement item domains (e.g., decisions regarding new cars, financial services, and apartments), the preferred preference aggregation strategies \cite{Masthoff2011} differ from low-involvement item domains such as restaurants and movies. The latter are often the domains of repeated group decisions (e.g., the same group selects a restaurant for a dinner every three months). Groups tend to apply strategies such as \emph{Least Misery} (LMS), in high involvement item domains, and to prefer \emph{Average Voting} (AVG) in low-involvement item domains.  When recommending packages, the task is to recommend a set of items in such a way that individual group members perceive the recommendation as fair \cite{Serbos2017}. One interpretation of fairness stated in Serbos et al. \cite{Serbos2017} is that there are at least $m$ items included in the package that a group member likes. 

An approach to take into account fairness in \emph{repeated group decisions} is presented by Quijano-Sanchez et al. \cite{Sanchez2013SocialFactors}, where rating predictions are adapted to achieve fairness in future recommendation settings. This adaptation also depends on the personality of a group member. For example, a group member with a strong personality who was treated less favorably last time, will be immediately compensated in the upcoming group decision. A similar interpretation of fairness is introduced in Stettinger  et al. \cite{Stettinger2014RecSysDC} where fairness is also defined in the context of repeated group decisions, i.e., decisions that repeatedly take place within the same or stable groups (groups with a low fluctuation). Fairness in this context is achieved by introducing functions that systematically adapt  preference weights, i.e., group members whose preferences were disregarded recently receive higher preference weights in upcoming decisions. For example, in the context of repeated decisions (made by the same group) regarding a restaurant for a dinner, the preferences of some group members are more often taken into account than the preferences of others. In such scenarios, the preference weights of individual group members can be adapted \cite{Stettinger2014RecSysDC} (see Formulae \ref{importanceadaptationfairness}--\ref{fairnessfunction}). 

Formula \ref{fairnessfunction} provides a  \emph{fairness} estimate per user $u_i$ in terms of the share of the number of supported preferences in relation to the number of defined preferences. The lower the value, the less the preferences of a user (group member of group $G$) have been taken into account, and the lower the corresponding degree of fairness with regard to $u_i$. Formula \ref{importanceadaptationfairness} reflects an approach to increasing fairness in upcoming recommendation sessions. If the fairness (Formula \ref{fairnessfunction}) in previous sessions was lower than average, a corresponding upgrade of user-specific importance weights ($w$) takes place for each dimension. For an example of adapted weights see Table \ref{tab:FairnessInGroupDecisionMaking:6}. 

\begin{equation} \label{importanceadaptationfairness}
w'(u_i,dim_j) = w(u_i,dim_j) \times (1+(\frac{\Sigma_{u \in G}fair(u)}{|G|}-fair(u_i)))
\end{equation}

\begin{equation} \label{fairnessfunction}
fair(u_i) = \frac{\#supported preferences(u_i)}{\#group~decisions}
\end{equation}


\begin{table*}[ht!]
\center
\begin{tabular}{|C{1.0cm}|C{0.75cm}|C{0.75cm}|C{0.75cm}|C{1.5cm}|C{0.75cm}|C{0.75cm}|C{1.0cm}|} 
\hline
user 	& \multicolumn{3}{  c  |}{importance (imp)}	&  fairness (fair) 	& \multicolumn{3}{  c  |}{adapted importance (imp')} \\ \cline{2-4} \cline{6-8}
  		& $dim_1$	&  $dim_2$ & $dim_3$ 		&   			& $dim_1$	&  $dim_2$ 	& $dim_3$ 		\\ \hline
$u_1$ 	& 0.3		&   0.3	 	& 0.4			&  4/8=0.5 		& 0.375		&  0.375    &  0.5			\\ \hline 
$u_2$  	& 0.5		&   0.4	 	& 0.1			&  6/8=0.75 	& 0.5		&  0.4    	&  0.1			\\ \hline 
$u_3$  	& 0.3		&   0.2	 	& 0.5			&  8/8=1.0 		& 0.225		&  0.15    	&  0.375		\\ \hline 
\end{tabular} \vspace{0.2cm}
\caption{An example of an adaptation of individual users' weights to take \emph{fairness} into account. In this example, the importance ($imp$) weights of user $u_1$ have been increased, the weights of $u_2$ remain the same, and the weights of user $u_3$ have been decreased (the preferences of $u_3$ have been favored in previous decisions -- a visualization is given in Figure \ref{fig:Fairness:6}).}
\label{tab:FairnessInGroupDecisionMaking:6}
\end{table*}

\index{explanations ! visualization} \index{visualization ! constraint-based recommendation} \index{constraint-based recommendation} \subsubsection*{Visualization of Constraint-based Explanations for Groups} 
\begin{figure}[ht!]
\centering
\includegraphics[scale=.4]{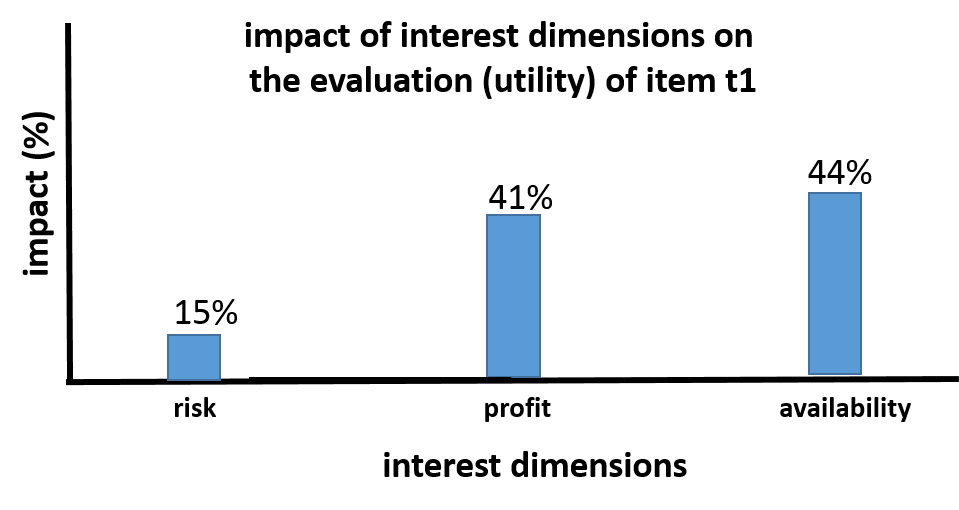}
\caption{Visualization of the importance of interest dimensions with regard to the overall item evaluation (the importance values are based on Table \ref{tab:CBRExplanationsPromotionAggregatedPredictionsMAUT:6} where $dim_1=risk$, $dim_2=profit$, and $dim_3=availability$).}
\label{fig:utilitymautexplanations:6}       
\end{figure}

An example of visualizing the importance of interest dimensions with regard to a final evaluation (utility) is given in Figure \ref{fig:utilitymautexplanations:6}.  Examples of interest dimensions when evaluating, for example, financial services, are \emph{risk}, \emph{profit}, and \emph{availability}.

\begin{figure}[ht!]
\centering
\includegraphics[scale=.45]{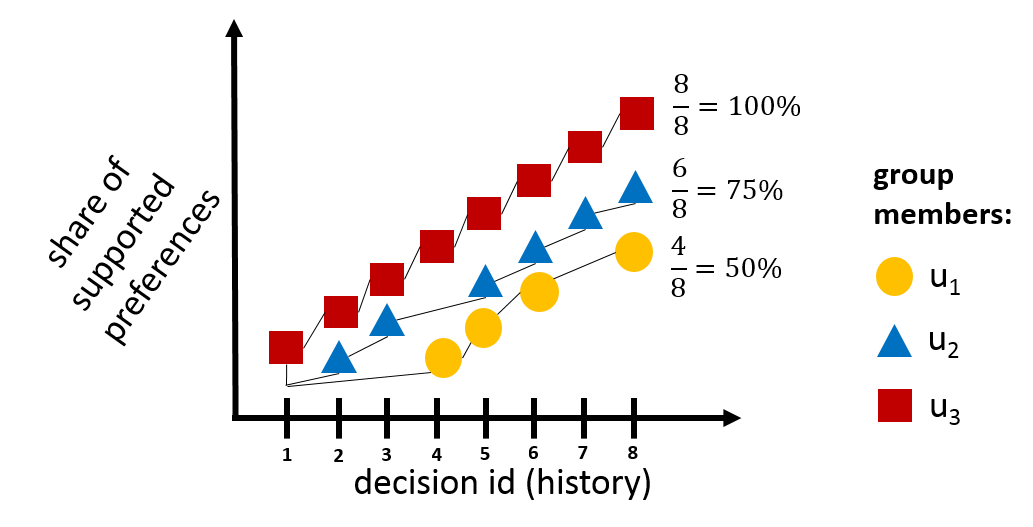}
\caption{Visualizing the \emph{degree of fairness} (Formula \ref{fairnessfunction}) in repeated group decisions (e.g., decisions on restaurant visits). In this example, the visualization indicates that user $u_1$ was at a disadvantage in previous decisions.}
\label{fig:Fairness:6}       
\end{figure}

If the degree of fairness of previous group decisions has to be made transparent to the group, for example, for explaining adaptations regarding the importance weights of individual group members, this can be achieved on the basis of a visualization as depicted in Figure \ref{fig:Fairness:6}. An example of a related verbal explanation is the following: '\emph{the interest dimensions favored by user $u_1$ have been given more consideration since $u_1$ was at a disadvantage in previous decisions}'.

\section{Critiquing-based Recommendation}\label{sec:cbcrtf:6}

\index{explanations ! critiquing-based recommendation} \index{critiquing-based recommendation} To assist users in constructing and refining preferences, critiquing-based recommender systems \cite{Chen2012} determine recommendations based on the similarity between candidate and reference items. For example, in the domain of digital cameras, related explanations  focus on item attributes such as \emph{price}, \emph{resolution}, and \emph{optical zoom}. \emph{System-generated critiques} (e.g., compound critiques \cite{McCarthy2005}) help to explain the relationship between the currently shown reference item and candidate items. Such explanations have been found to help  educate users and increase their \emph{trust} in the underlying recommender system \cite{PuChen2007}. 

\subsubsection*{Critiquing-based Explanations for Groups}  \emph{User-defined critiques}, i.e., critiques on the current reference item directly defined by the user, can be used for the generation of explanations for recommended items (see the example in Table \ref{tab:CBRExplanationsPromotionAggregatedProfilesRequirements:6}).  

\begin{table*}[ht!]
\center
\begin{tabular}{|C{1.75cm}|C{1.0cm}|C{1.0cm}|C{1.0cm}|C{1.25cm}|C{1.25cm}|C{1.25cm}|C{1.25cm}|} 
\hline
\multicolumn{4}{|  c  |}{critiques of group members} &  \multicolumn{3}{  c  |}{support(attribute,$t_i$)}  			\\ \cline{1-7}
attribute  		&  crit($u_1$)  &   crit($u_2$)	&   crit($u_3$) &  $t_1$ 		&  $t_2$ 			&  $t_3$ 			\\ \hline
price 			&  $\leq$1.000	&  $\leq$750  &   $\leq$600	&  299(1.0)		&	650	(0.66)		&   1.200 (0.0)		\\ \hline 
res  			&  $\geq$20		&  $\geq$18   &   $\geq$25  &  24(0.66)		&	25	(1.0)		&   30 (1.0)		\\ \hline 
weight  			& $\leq$1		&  $\leq$2	  &   $\leq$1	&  1.5(0.33)	&   3 (0.0)			&   2 (0.33)		\\ \hline 
exchangeable lens   &  y        &  y            &   n			&  y(0.66)		&	y (0.66)		&   n (0.33)		\\ \hline 
\end{tabular} \vspace{0.2cm}
\caption{Critiques of group members as a basis for generating explanations for item recommendations. \emph{Support} is defined by the share of attribute-specific critiques supported by an item $t_i$.}
\label{tab:CBRExplanationsPromotionAggregatedProfilesRequirements:6}
\end{table*}

In this context, $support(attribute,t_i)$ (see Formula \ref{supportcritiquingfunction}) indicates how often an item supports a user critique on the \emph{attribute}. For example, item $t_1$ supports a critique on \emph{price} three times since all the critiques on \emph{price} are consistent with the price of $t_1$, i.e., \emph{support}($price,t_1$)=1.0. However, \emph{support}($weight,t_1$) is only $0.33$ since the weight of $t_1$ is $1.5$ which is inconsistent with two related critiques.

\begin{equation} \label{supportcritiquingfunction}
support(attribute, t_i) = \frac{\# supportedcritiques(attribute,t_i)}{\#critiques(attribute)}
\end{equation}

On the verbal level, an explanation for item $t_1$ could be '\emph{the price of camera $t_1$ ($299$) is clearly within the limits specified by the group members. As expected, it has an exchangeable lens. It has a resolution (24) that satisfies the requirements of $u_1$ and $u_2$, however, $u_3$ has to accept minor drawbacks. Furthermore, the weight of the camera (1.5) is significantly higher than expected by $u_1$ and $u_3$}'. 

Such explanations can be provided if the preferences of group members are known. Otherwise, explanations have to be generated on the basis of \emph{aggregated models}, where item properties are compared with the aggregated critiques defined in the group profile.

\index{visualization ! critiquing-based recommendation} \index{visualization} \subsubsection*{Visualization of Critiquing-based Explanations for Groups} An example of visualizing the support of different attribute-specific critiques is given in Table \ref{tab:visualizationcritiquingbasedrecommendation:6}.  The $\surd$ symbol denotes the fact that the user critique on an attribute of item $t_i$ is supported by $t_i$.

\begin{table*}[ht!]
\center
\begin{tabular}{|C{1.5cm}|C{1.5cm}|C{2.0cm}|C{2.5cm}|C{3.5cm}|} 
\hline
user 		& \multicolumn{4}{  c  |}{attributes($t_1$)}	\\ \cline{2-5}
  			& $price=299$	&  $resolution=24$ & $weight=1.5$ 	& $exchangeable lens=y$	\\ \hline
$u_1$ 		& $\surd$	&   $\surd$	 	& $\times$	& $\surd$				\\ \hline 
$u_2$  		& $\surd$	&   $\surd$	 	& $\surd$	& $\surd$				\\ \hline 
$u_3$  		& $\surd$	&   $\times$	& $\times$	& $\times$				\\ \hline 
\end{tabular} \vspace{0.2cm}
\caption{Summarization of the support-degree of user-specific critiques on item $t_1$.}
\label{tab:visualizationcritiquingbasedrecommendation:6}
\end{table*}

\section{Conclusions and Research Issues}\label{sec:conclusions:6}

In this paper, we provided an overview of explanations that help single users and groups to better understand item recommendations. As has been pointed out in pioneering work by Jameson and Smyth \cite{JamesonSmyth2007}, explanations play a crucial role in group recommendation scenarios. We discussed possibilities of explaining recommendations in the context of the basic recommendation paradigms of collaborative filtering, content-based filtering, constraint-based, and critiquing-based recommendation, taking into account specific aspects of group recommendation scenarios. In order to support a more in-depth understanding of how explanations can be determined, we provided a couple of working examples of verbal explanations and corresponding visualizations. 

Although extensively analyzed in the context of single-user recommendations (see, e.g., Tintarev  \cite{Tintarev2009}), the generation of explanations for groups entails a couple of open research issues.  Specifically, aspects of group dynamics have to be analyzed with regard to their role in generating explanations. For example, \emph{consensus}, \emph{fairness}, and \emph{privacy} are major aspects -- the related research question is how to define explanations that best help to achieve these goals. Some initial approaches exist to explain the application of aggregation functions in group recommendation contexts (see, e.g., Ntoutsi et al. \cite{Ntoutsi2012}), however, a more in-depth integration of social choice theories into the generation of explanations has to be performed. This is also true on the algorithmic level, as in the context of group-based configuration \cite{FelfernigOpenConfiguration2014}. In this context, the integration of information about personality and emotion into explanations has to be analyzed.  Initial related work can be found, for example, in Quijano-Sanchez et al. \cite{Sanchez2017} where social factors in groups are taken into account to generate \emph{tactful explanations}, i.e., explanations that avoid, for example, damaging friendships. 

Mechanisms that help to increase the quality of group decision making processes have to be investigated \cite{Konstan2012}. For example, explanations could also be used to trigger intended behavior in group decision making such as exchange of decision-relevant information among group members \cite{Atas2017}. Finally, explaining hybrid recommendations \cite{Kouki2017} and recommendations generated by matrix factorization (MF) approaches \cite{Abdollahi2017,Rastegarpanah2017} are issues for future research. Summarizing, explanations for groups is a highly relevant research area with many open issues for future work. 


\bibliographystyle{IEEEtran}
\bibliography{IEEEabrv,bibliography}

\end{document}